\documentclass[12pt,a4paper]{article}

\usepackage{amsmath,amssymb,amsthm,graphicx,color,bm,soul}
\usepackage{hyperref}
\usepackage{leftidx}

\begin{document}


\title{Constructing a bridge between functioning of oscillatory neuronal networks and quantum-like cognition along with quantum-inspired computation and AI}

\author{Andrei Khrennikov$^{1,*}$ Atsushi Iriki$^{1}$, and  Irina Basieva$^{1}$}
\date{}                     
\maketitle
\noindent
$^{1}$Center for Mathematical Modeling 
in Physics and Cognitive Sciences\\
Linnaeus University, V\"axj\"o, SE-351 95, Sweden\\
$^{2}$Teiko University Advanced Comprehensive Research Organization\\
Division of Artificial Intelligence\\
2-21-1 Kaga, Itabashi-ku, Tokyo 173-0003, Japan

*Corresponding author email: Andrei.Khrennikov@lnu.se

\date{}                     

\maketitle

\abstract{Quantum-like (QL) modeling, one of the outcomes of the quantum information revolution, extends quantum theory methods beyond physics to decision theory and cognitive psychology. While effective in explaining paradoxes in decision making  and 
effects in cognitive psychology, such as conjunction, disjunction, order, and response replicability, it lacks a direct link to neural information processing in the brain. This study bridges neurophysiology, neuropsychology, and cognitive psychology, exploring how oscillatory neuronal networks give rise to QL behaviors.  Inspired by the computational power of neuronal oscillations and quantum-inspired computation (QIC), we propose a quantum-theoretical framework for coupling of cognition/decision making and neural oscillations - {\it QL oscillatory cognition.} This is a step, may be very small, towards clarification of the relation between mind and matter and the nature of perception and cognition. We formulate four conjectures within QL oscillatory cognition and in principle they can checkedAsanoexperimentally. But such experimental tests need further theoretical and experimental elaboration. One of the conjectures (Conjecture 4) is on resolution of the binding problem by exploring QL states entanglement generated by the oscillations in a few neuronal networks. Our findings suggest that fundamental cognitive processes align with quantum principles, implying that humanoid AI should process information using quantum-theoretic laws. Quantum-Like AI (QLAI) can be efficiently realized via oscillatory networks performing QIC. }   

\section{Introduction}

A significant outcome of the quantum information revolution (known as second quantum revolution)  is the extension of quantum theory into cognitive studies, specifically those relevant to  decision-making (option selection) \cite{Zeit}. This broad research area, known as {\it quantum-like} (QL) modeling (QLM), spans fields such as molecular biology,genetics, epigenetics, evolution biology, finance, economics, and social and political science.\footnote{See monographs \cite{QL0,UB_KHR,Busemeyer,Haven,QL3,Bagarello,QLH,Open_KHR, Bagarello1}, the Palgrave Handbook \cite{handbook}, and recent reviews \cite{Pothos,Khrennikov}.} Importantly, QLM should be distinguished from quantum biophysics \cite{QBIOP,Melkikh1,Melkikh2} and, specifically, from quantum cognition and consciousness \cite{H,P,V1,V2,V3,Igamberdiev1,Igamberdiev2,Igamberdiev2a}. While quantum biophysics and quantum cognition deal directly with quantum physical processes in biological systems, QLM focuses on macroscopic biological, social, and, more recently, AI systems. This approach describes information processing using quantum information and probability principles, though it is not rooted in quantum physics. QLM has been successfully applied to reinterpret classical decision theory paradoxes (e.g., the Ellsberg paradox) and to model key cognitive psychology phenomena such as conjunction, disjunction, order, and response replicability effects, and contextuality.\footnote{See, e.g., 
\cite{Aerts,pMENTAL,Khrennikov2002,KHC3,KHR_IM,Aequbitts0,ASR}, \cite{Behti,BioBas,Bruza,Bruza2,B0,SR,Bagarello,Bagarello1,Bagarello2},
\cite{Kvam,Pothos23,Gallus}, \cite{Wang,Wang1,Wi},  \cite{Iriki1,PLOS,TC}, \cite{Gunji1, Gunji2}, \cite{OzawaE,OzawaJMP}}. QLM has also been applied in molecular biology, genetics and epigienetics - genomes, cells, as well as genetic and cellular networks are also 
decision makers  working on option selections and demonstrating complex behavior \cite{Asano0,Asano,QIB,QL3,AsanoD}.   
More recently, the QL approach has begun to be used in medical diagnostics, particularly for neurological disorders \cite{Shor1}-\cite{Shor2}. In these works, the quantum(-like) potential of Bohmian pilot wave theory constructed on the basis of classical EEG signals was used as a bio-information marker of 
neurological disorders (and this QL-machinery work very well).     

In QL studies, quantum theory serves as a phenomenological framework. Mental states, such as the beliefs of a decision-maker, are represented as quantum states — either normalized vectors in a complex Hilbert space or, more generally, density operators. Cognitive observables, including questions, tasks, or traits in biology, are represented as Hermitian operators or, more broadly, as quantum instruments \cite{BioBas,OzawaE,OzawaJMP}. This representation provides clarity on certain aspects of human cognition  relevant  to decision-making 
(option selection) that classical probability and information theory struggle to explain.

However, as previously mentioned, QL theory remains a phenomenological framework. From a neurophysiological perspective, it is unclear why QL theory aligns so well with cognitive phenomena that are not directly tied to quantum physics in the brain or body. For example, 
``genuine quantum cognition and consciousness'' (see, e.g., \cite{H,P,V1,V2,V3,Igamberdiev1,Igamberdiev2,Igamberdiev2a})
 explores  the collapse of wavefunctions in quantum theory. In contrast, QL theory does not rely on such a complex and potentially ambiguous concept as wavefunction collapse. Furthermore, when representing observables using Hermitian operators, QL studies face a more intricate situation than in quantum physics. In quantum theory, the quantization process generates quantum operators from classical phase-space variables (e.g., Schr\"odinger quantization). However, in cognitive science, there is no cognitive analog of classical phase-space mechanics, which means there is nothing to quantize in this context.

QL studies really cry for establishing connection between the neuronal and quantum representations of cognition. Some steps in this direction were done in articles \cite{DBS,DB,TC,BUSN,Scholes1,Scholes2} (see also \cite{Bagarello2,Wi}. The present work is a natural continuation of article \cite{GPT} in that the {\it generalized probability theory}  \cite{Holik} (operational measurement theory \cite{Davies-Lewis,DV,Ozawa1}) was employed for construction of a bridge between functioning of neuronal networks in the brain and the generalized quantum formalism for cognition relevant to  decision making.
But the formalism used in \cite{GPT} differs from the formalism based on complex Hilbert space. The aim of this paper is to show that brain's functioning at the neuronal level can be represented within the standard quantum formalism.

This complex Hilbert space representation is inspired by the discussion in the article \cite{Wolf} that compares the brain's functioning to information processing in QIC devices (analog and digital) discussed in section \ref{QICS} (see also \cite{Iriki1}). These devices utilize harmonic oscillators for the classical realization of qubit-like information units \cite{Goto0,Goto}. 
As noted in \cite{Wolf}, ``biological neuronal networks, particularly those found in the cerebral cortex and hippocampus, also exhibit oscillatory dynamics.'' The computational performance of such dynamics can be examined. 
 And ``it was found  that configuring network nodes as damped harmonic oscillators rather than leaky integrators greatly enhanced   performance, sometimes by orders of magnitude, relative to known RNNs'' (recurrence neuronal networks). In \cite{Wolf} this over-performance was coupled to the QIC features of biological systems 
(see also \cite{Singer1,Singer2,SingerG1,SingerG2}).

In this paper, we construct the quantum-like representation for networks of harmonic oscillators. This approach has several foundational implications:

\begin{itemize} 
\item It links QL cognition with neuronal information processing in the brain. 
\item It supports the study in \cite{Wolf} on the enhanced computational ability of oscillatory neuronal networks in the brain, particularly in the cerebral cortex and hippocampus.
\item It strengthens the case for the supremacy of QIC. 
\item It fosters the integration of QL cognition, oscillatory neuronal computation in the brain, and QIC with humanoid AI. 
\item It provides a neuronal justification for applying QL methods in the medical diagnosis of neurological disorders.
\end{itemize}
It is important to note that QIC is not the primary focus of this article; it is only discussed in section \ref{QICS} in relation to the studies in \cite{Wolf}. The coupling of humanoid AI with quantum theory is addressed in section \ref{AI}. This represents the first step toward the development of QL-based Artificial Intelligence (QLAI) exploring QIC with oscillatory networks.

Following \cite{SingerG2}, we assume that an oscillatory {\it ``node in neuronal networks should not be considered to represent a single biological neuron, but rather as a microcircuit composed of recurrently coupled populations of excitatory (E) and inhibitory (I) neurons ...'' }In this sense, an oscillatory network under consideration {\it ``should be understood 
as a mesoscale model of a cortical network.''}

The QL representation of the functioning of oscillatory networks is based on the formalism of {\it prequantum classical statistical field theory} (PCSFT) \cite{Beyond,annals}, in which random fields serve as hidden variables for the quantum formalism. In this framework, the covariance operators of these random fields, after proper normalization, are treated as density operators. QL observables, represented by Hermitian operators, emerge as the images of the quadratic forms of the ``subquantum random fields.'' Some attempts to apply this formalism to cognition have been made in \cite{Khrennikov2010}, where electromagnetic fields in the brain were considered to perform cognitive information processing. Building on insights from \cite{Wolf} (and \cite{Singer1, Singer2, SingerG1, SingerG2}), we now focus on the discrete sources of random oscillations (ROs) generated in neuronal circuits. We apply the PCSFT methodology to construct the QL representation of the following:

\begin{itemize} 
\item Density operators, QL states, as (proper normalized)  covariance operators of ROs , 
\item Hermitian operators, QL observables, as the quadratic forms of ROs, 
\item QL averages, of the form $\langle A \rangle_\rho = \mathrm{Tr} (\rho A)$ as the averages of quadratic forms of ROs. 
\end{itemize}
In the QL model, density operators correspond to cognitive (mental) states. This establishes a coupling between the correlations within the neuronal network $S$ and the cognitive state it generates. Thus, cognition in the QL model is associated with the correlations between oscillations in the nodes of $S$ (see section \ref{cognitive2} for further discussion). The strength of these correlations can serve as a quantitative measure of cognition ``generated'' by $S.$

For a compound neuronal network $S=(S_1,S_2),$ its QL state is coupled  with the covariance operator of ROs, but coupling is mathematically more complicated than for a single system (section \ref{entanglement}). In particular, to generate entangled states ROs in $S$ should be accompanied with a background field. In quantum physics, this is the ``zero point field'' - the electromagnetic field of vacuum fluctuations. Int QL oscillatory cognition the candidates for such background fields are the electromagnetic fields corresponding to the basic rhythms in the brain.   

QL oscillatory cognition suggests the resolution of {\it the binding problem:} an object, its background, and symbolic or emotional characteristics are combined into a single experience via generation of QL entangled state by oscillatory networks interacting with the background electromagnetic field.

Additionally, we refer to the articles \cite{Scholes1,Scholes2}, in which the QL representation is used to describe information processing in neuronal or molecular networks. These articles primarily rely on graph theory and the structure of physical connections between the network's nodes, such as axon-dendrite connections between neurons in the brain. In contrast, in our framework, as well as in the article \cite{GPT}, the graph structure of networks is embedded in covariance operators, though this occurs only indirectly.

Section \ref{DS} discusses the coupling of neuronal circuit network dynamics with quantum dynamics for isolated systems (the Schr\"odinger equation) and open systems, including the {\it Gorini-Kossakowski-Sudarshan-Lindblad} (GKSL) equation \cite{Open_KHR} and more general (nonlinear) master equations for the density operator (see \cite{KHRBNL}, cf. \cite{Ottinger1,Ottinger2,Tsekov}).\footnote{ We emphasize that the quantum theoretical representation of classical Hamiltonian systems (section \ref{DS}) should be sharply distinguished from the standard quantization procedure (e.g., Schr\"odinger quantization in phase space). Our approach is historically linked to the {\it {\it Riemann-Silberstein representation} for the classical electromagnetic field}, $Z(t,x) =E(t,x) + i B(t, x),$ where $E$ and $B$ are the electric and magnetic vector-components of the field.
 In this representation, Maxwell's equations are written in the form of the Schr\"odinger equation (section \ref{appendix A}). Similarly, the dynamics of a system of coupled harmonic oscillators can be represented as a Schr\"odinger equation (section \ref{DS1}). In section \ref{Gnetwork}, the linear dynamics of the covariance matrix of ROs in neuronal circuit networks is represented by a generally nonlinear master equation for the corresponding QL state - density operator.} In section \ref{appendix B}, we explore the QL representation for systems of coupled damped harmonic oscillators. This topic is important for establishing a connection with the cognitive-computational model presented in \cite{Wolf}.

This article represents a small, yet important, step toward clarifying the relationship between mind and matter, as well as the nature of perception and cognition (cf. with similar line of thinking \cite{Singer3, Iriki0,Iriki}). Motivated by extensive research on the connection between the functioning of oscillatory neuronal networks in the brain and cognition and consciousness \cite{Singer1, Singer2, SingerG1, SingerG2}, we have developed a bridge between the correlations in these oscillatory networks and the quantum-like (QL) model of {\it cognition/decision making.} This model captures several unique (``non-classical'') properties of cognition.

 We come with three conjectures on QL features of brain's functioning, Conjecture 1 (section \ref{QFDM}),  Conjecture 2 
(section \ref{cognitive1}), Conjecture 3 (section \ref{QLO}). The big challenge is design experimental procedures to verify these conjectures.

\section{Quantum formalism for decision making}
\label{QFDM}

As the shortest introduction to quantum-like modeling, we briefly present the scheme of decision-making formalized in quantum-theoretical framework. Let ${\cal H}$  be a complex Hilbert space with the scalar product $\langle \cdot|\cdot\rangle$ and the corresponding norm $||\psi||^2= \langle \psi|\psi\rangle.$ Denote the set of density operators in ${\cal H}$ by the symbol 
${\cal D}= {\cal D}({\cal H}).$ These are positive Hermitian trace one operators.   

The QL  states of a cognitive system, decision maker, are represented by normalized vectors of a complex Hilbert space 
${\cal H}$ or more generally by density operators. So, ${\cal D}$ is the space of QL states. QL observables -
questions, tasks, traits - are represented by Hermitian operators, $A:  {\cal H} \to {\cal H}.$ The spectrum of $A$ quantifies possible outcomes of observations, cognitive measurements. Typically in QLM   ${\cal H}$ is finite dimensional and the spectrum of $A$ is discrete and consist of eigenvalues, $x_1,...,x_n$  labeling the outcomes of observations, e.g., the possible answers to question $A.$
In the finite dimensional case one can operate with matrices.
By Born's rule the average of $A$ is given by the formula
\begin{equation}
\label{DM}   
\langle A\rangle_\rho = \rm{TR} \rho A.
\end{equation}
One of the main distinguishing features of this rule is its linearity w.r.t. the state and observable (similarly to 
classical Kolmogorov probability theory \cite{K} in that averages are given by integrals w.r.t. probability measures). 

As was mentioned, there is no quantization rule and QL states and observables are constructed ``by hands'' to design a model matching statistical data (e.g., from cognitive psychology or social science) with formula (\ref{DM}).  

The QL representation is successfully explored in cognitive psychology and decision making. We theorize that  

\medskip

{\bf Conjecture 1.} {\it  The brain is able to operate within QL representation that is by using the QL matrix calculus for information processing.}

\medskip

QL approach to cognition can be treated as a part  Emergentism (see \cite{Kriegel} for a review; cf. \cite{Wolf}).  (Emergentist  theories assume that the mental or spiritual dimension is an emergent property of but not identical with brain processes.)


\section{Quantum-like representation of oscillatory networks}

Consider a network $S$ that nodes are oscillators coupled to each other.  Such node represent a neuronal microcircuit composed of recurrently coupled populations of neurons. Such oscillatory neuronal circuits are the elementary information processing units. The term ``coupling'' doesn't mean solely connections via axons and dendrites. For electromagnetic oscillators (section \ref{appendix A}), coupling can be performed via electromagnetic fields.     

 In the simplest (but very general) model, the  dynamics of oscillations in each node is described by two variables, 
$q_j=q_j(t)$ and $p_j(t)=\dot q_j(t), j=1,...,N,$ where $N$ is the number of elementary circuits, oscillators, in $S.$  
Set $Q= \mathbb{R}^N, P=\mathbb{R}^N$ and $X=Q\times P= \mathbb{R}^{2N},$ the classical phase space of the system - classical oscillatory phase space. The classical  state of the network  $S$ is a point $x=(q,p)$ of the phase space,  here $q=(q_1,...,q_N)$ and $p=(p_1,...,p_N).$ The phase space $X$ is endowed with scalar product, $(x,y)= (q_x,q_y) + (p_x,p_y)= \sum_j (q_{xj}q_{yj} + p_{xj} 
p_{yj}).$   

To proceed to QL representation in complex Hilbert space, we introduce the complex variables  $z_j= (x_j+i p_j)/\sqrt{2}.$ Thus, for each instant of time,  the network state is described as the complex vector $z=(z_1,..., z_N).$\footnote{By making the variable $j$ continuous, as $x \in \mathbb{R}^3,$ we consider electromagnetic oscillators with 
$q(x)=E(x)$ and $p(x)= B(x),$ where $E(x)$ and $B(x)$ are electric and magnetic components of the field. Then 
$Z(x)= E(x) + i B(x)$ is the Riemann-Silberstein  representation of the electromagnetic field (see section \ref{appendix A} for the details).
} This description  corresponds to complexification of the phase space $X=Q\times P$  and exploring the complex Hilbert space  
${\cal H}= Q\oplus  i P= \mathbb{C}^N.$ It is endowed with the scalar product 
$\langle u| v \rangle = \sum_{j} u_j \bar{v}_j$ and the corresponding norm  
$||z||^2 = \langle z| z \rangle = \sum_{j} |z_j|^2.$ The network nodes determine the canonical orthonormal basis $(e_j=|j\rangle, 
j=1,2,...,N,$ in ${\cal H}.$ Symplectic linear transformations  in phase space $X$ correspond to ($\mathbb{C}$-)linear transformations in ${\cal H}.$   

We speculated (in section \ref{QFDM}) that the brain can explore the QL representation in complex state space. So, it can perform transitions from one orthonormal basis in  ${\cal H}$ to another. In particular, the transition to the normal coordinates for a system of coupled harmonic oscillators corresponds to the transition to the special basis in QL state space ${\cal H}.$  Such basis is generally ``nonlocal'', since its vectors are linear combinations of node-vectors $|j\rangle.$ The frequencies corresponding to normal coordinates are eigen-frequencies
of neuronal network $S.$  

The multiplex structure of a network $S$ and complexity of the signal exchange within $S$ and with its neuronal and physicochemical environment leads to the description of the  complex oscillatory variables  as random variables; the dynamics of the state of  each node is mathematically described by a stochastic process $z_j= z_j(t; \omega),$ where $\omega$ denotes a random parameter. We emphasize that this is classical randomness (not so called quantum randomness \cite{VN}).    For the moment, we consider $S$'s state for the fixed instance of time $t$ and omit this variable, i.e., the state is given by the random complex vector $z=z(\omega);$ the dynamics will be studied in section \ref{DS}.  

\subsection{Quantum-like cognitive states as covariance operators of classical oscillatory states}
\label{cognitive1}

The QL state of $S$  is given by the (complex) covariance matrix of the classical state given  the random complex vector $z:$ 
\begin{equation}
\label{lll}
C= (c_{km}), \; c_{km}=E [z_k \bar{z}_m] = \int_\Omega z_k(\omega) \bar{z}_m(\omega) dP (\omega), 
\end{equation}
where $P=P_S$ is the probability measure describing randomness inside network $S$ and $\Omega$ is the set of random parameters.
The  covariance  matrix $C$ is 
\begin{itemize}
\item Hermitian that is $C^\star= C,$ i.e., $c_{ij}= \bar{c}_{ji};$   
\item positively defined $C \geq 0$ that is $\langle z|C| z\rangle \geq 0$ for any $z \in \mathbb{C}^N.$ 
\end{itemize}
We note that density matrices of quantum theory are Hermitian and positively defined as well.
Let us normalize the covariance matrix $C$ by setting 
\begin{equation}
\label{lll1}
C \to  \rho= F_s(C)= C/ \rm{Tr} C   
\end{equation}
(the index $s$ is for ``state'').
Then $\rho$ has trace one and it is a density matrix.
Thus, the random classical states of the network $S$ of random oscillators can be mapped to quantum states. The map $F_s$ isn't 
one-to-one, scaling of the covariance matrix $C$ with a positive factor $c$ leads to the same QL state $\rho.$

We note that 
\begin{equation}
\label{D}
\rm{Tr} C = E[||z||^2]= \int_\Omega ||z||^2 dP (\omega).   
\end{equation}
If all oscillators have zero mean values, this is simply the dispersion of the ROs expressed in the complex variables.
Thus, in (\ref{lll1})  the covariance matrix is normalized by the spread of random oscillatory variables. 

We theorize on the biological meaning of density operators in QLM:   

\medskip

{\bf Conjecture 2.}  {\it The belief states (mental states) considered in QLM of cognition relevant to decision making are 
coupled to neurophysiological processes in the brain via correspondence given by (\ref{lll1}). In the brain our believes are represented as the correlations between oscillating neuronal circuits forming circuits' networks} (cf. \cite{Wolf,Singer1,Singer2,SingerG1,SingerG2}).  

\medskip

Generally a covariance matrix doesn't determine a random vector uniquely. If we restrict consideration to jointly Gaussian 
random vectors, then the real covariance matrix determines uniquely the Gaussian random vector - up to the mean value. However, this is not valid for complex covariance matrices that are under consideration. For them, to determine uniquely the jointly Gaussian random vector, one should restrict consideration to circularly symmetric random variables.\footnote{A random variable $z$ is circularly symmetric if, for any angle $\theta,$ the random variables $z$ and $e^{i \theta} z$ have the same probability distribution.
We note that such random variables have zero mean value. The condition of invariance of the probability distribution
w.r.t. scaling w.r.t. phase factor $e^{i \theta}$ is similar to the condition of determination 
of pure quantum states up to such factor.} 

 Thus, the QL representation gives a fuzzy picture of what is happened at the level of ROs in neuronal networks. Hidden variables exist (as points of the oscillatory phase space), but they are invisible in the QL representation. We stress that we consider the QL representation of a single system $S.$ For a compound system $S$ of say two networks, $S=(S_1,S_2),$ the problem of coupling of 
between classical oscillatory and QL representations is more complicated (section \ref{entanglement}).   

We further proceed in the operator framework and speak not about covariance and density matrices, but about operators acting 
in a complex Hilbert space ${\cal H}$  
For simplicity, we assume that   ${\cal H}$ is finite dimensional  (besides section \ref{appendix A} devoted to electromagnetic oscillators).
Let $(e_j)$ be an orthonormal basis in ${\cal H}.$ Then  any $z\in {\cal H}$ can be expanded w.r.t. this basis as
$z= \sum_j z_j e_j,$ where $ z_j=\langle z|e_j\rangle.$ The covariance operator $C$ of a Hilbert space valued random variable 
$z=z(\omega)$ can be defined by its bilinear form ($h_1,h_2 \in{\cal H}):$
\begin{equation}
\label{average0}
\langle C h_1|h_2\rangle= E[ \langle h_1|z \rangle \langle z|h_2\rangle]=
\sum_{jk} \langle h_1|e_j \rangle \langle e_k|h_2 \rangle E[\langle e_j|z \rangle \langle z|e_k \rangle]=
\end{equation}
$$
\sum_{jk}  E [z_k \bar{z}_j] h_{1j} \bar{h}_{2k} = \sum_{jk}  c_{kj}  h_{1j} \bar{h}_{2k}. 
$$
Covariance operators are Hermitian and positively defined. 
Denote the set of such operators by the symbol ${\cal C}\equiv {\cal C}({\cal H}).$ 

By using the map (\ref{lll1}) the classical states of oscillatory random networks are represented by density operators acting 
in complex Hilbert ${\cal H}.$   Then 
\begin{equation}
\label{Dcor}
F_s: {\cal C} \to {\cal D}.
\end{equation}
              
\subsection{Pure states and their classical preimages}
\label{pure}

It is interesting to find the classical preimages of pure states. Consider deterministic oscillatory dynamics 
(see section \ref{DS}) that is $z(\omega)= z$ (a.e.), where $z$ is the fixed  vector belonging ${\cal H};$   the covariance operator is determined by the bilinear form $\langle C_z h_1|h_2\rangle= \langle h_1|z \rangle \langle z|h_2\rangle.$ Hence, $C_z= |z\rangle\langle z|.$  If we select in ${\cal H}$ an orthonormal basis, say the node-basis $(|j\rangle, j=1,...,N),$ i.e.,
$z= \sum_j c_j |j\rangle, c_j \in \mathbb{C},$ then $C_z=(c_{km}= z_k \bar{z}_m).$ Now by normalizing the operator $C_z$ by its 
trace which is equal $||z||^2,$ we obtain the density operator $\rho_\psi= F_s (C_z)$ corresponding to the pure state 
$|\psi\rangle = z/||z||.$  Thus, {\it the absence of ROs in a network $S$ is expressed as a pure quantum state.}   

Now consider an arbitrary discrete random variable $z=z(\omega)$ with the values $(z_1,...,z_m), z_j \in {\cal H},$ and 
$P(z=z_j)=p_j.$ Then
\begin{equation}
\label{mix}
C=\sum_j p_j C_{z_j}  =  \sum_j p_j |z_j\rangle\langle z_j|,
\end{equation}
Then $\rho= F_s(C)=  \sum_j p_j |\psi_j\rangle\langle \psi_j|,$ where  $|\psi_j\rangle= z_j/||z_j||$ is a pure state. This is the mixture representation of a density operator that is so useful in quantum theory. 

One may guess that pure QL states correspond to non-random oscillations in nodes of 
a neuronal network $S.$  But the situation is more complicated. Consider a circularly symmetric jointly Gaussian random vector
$\xi_z= \xi_z(\omega)$ with covariance operator $C_z= |z\rangle\langle z|.$  In the QL representation this Gaussian random vector  also corresponds to the pure state $|\psi\rangle,$ normalization of vector $z$ by its norm. (This situation reflects non-injectiveness of classical$\to$ quantum map (\ref{Dcor})). We remark that, although   $\xi_z$ is a random vector, its support is concentrated in one dimensional subspace 
${\cal H}_z= \{h= \lambda z, \lambda \in \mathbb{C}\}$ of ${\cal H}.$  Moreover, it is easy to show that any random vector $\xi=\xi(\omega)$ 
with the covariance operator $C_z$ takes its values in the subspace ${\cal H}_z$ with probability one. Let $h\perp {\cal H}_z.$ 
We have $0= \langle Ch| h \rangle= \int_\Omega |\langle h| \xi(\omega) \rangle|^2 dP (\omega)$ and, hence,  
$\langle h| \xi(\omega) \rangle=0$ (a.e.); this means that $\xi(\omega) \in {\cal H}_z$ (a.e.). 

As we have seen, a pure quantum state $|\psi\rangle$ represents ROs concentrated in the subspace 
${\cal H}_\psi.$ Since $|\psi\rangle$ is superposition of node-states $|j\rangle,$  the concentration of oscillations in 
one-dimensional subspace ${\cal H}_\psi$ is a ``nonlocal¨: oscillations are distributed over all nodes of 
the network. However, if the brain really employs the Hilbert space meta-representation, the concentration of ROs in one dimensional 
subspace is establishing a kind of order, QL order, for the network oscillations.      

\subsection{Quantitative measure of cognition}
\label{cognitive2}

In QLM quantum states, normalized vectors or generally density operators, represent cognitive (mental) states. 
This interpretation seems to be reasonable, since it works so well for numerous applications to cognition and psychology. 
The correspondence (\ref{Dcor}) transfers this interpretation from QL states to covariance operators of ROs.
Thus, the covariance operators of ROs also represent cognitive states generated by  neuronal networks. Hence, the covariance operator  $C_S$ of a neuronal network $S$ can be used for quantification of cognitive output of $S$-functioning. 

Consider the matrix $(c_{ij})$ of $C_S.$  It is natural to proceed with the matrix norm defined as
$||C||_{\rm{Frob}}^2= \sum_{ij} |c_{ij}|^2,$  {\it the Frobenius norm,} encountering all pairwise correlations between oscillations in network's nodes (including 
self-correlations of oscillations in each node). 
We can write this norm in the operator form. Denote by ${\cal L}={\cal L}({\cal H})$ the space of linear operators in 
${\cal H}.$ It is endowed with the scalar product $\langle A|B\rangle= \rm{Tr} A B^\star$ and the corresponding norm 
$||A||^2 = \langle A|A\rangle = \rm{Tr} A A^\star.$ Let $C$ be an arbitrary Hermitian operator. Then $||C||= ||C||_{\rm{Frob}}.$ The quantity 
\begin{equation}
\label{Dcogn}
\mu_{\rm{cogn}}= \rm{Tr} C^2= ||C||_{\rm{Frob}}^2
\end{equation}
is interpreted as a measure of oscillatory cognition.
Now we couple this quantity with quantum theory. Since $\rho_S= C_S/ \rm{Tr} C_S,$ we have $||\rho_S||_2^2= \sum_{ij} |\rho_{ij}|^2= \rm{Tr} \rho_S^2=  \rm{Tr} C_S^2/ (\rm{Tr} C_S)^2.$ 

We recall  the quantity $\rm{Tr} \rho^2$ is called state's {\it purity}, it equals 1 only for 
pure states. Let us consider another quantum-mechanical quantity that is {\it linear entropy} $E_L(\rho)= 1 - \rm{Tr} \rho^2.$ This is a measure of disorder in a system; the minimal disorder $E_L(\rho)=0$ corresponds to pure states. In the terms of the covariance operator of ROs in $S,$ we have that 
\begin{equation}
\label{LE}
E_L(\rho_S)= 1- \frac{\rm{Tr} C_S^2}{(\rm{Tr} C_S)^2}.
\end{equation}
Then $E_L(\rho_S)=0,$ the minimal disorder in $S,$ corresponds to concentration of oscillations in one dimensional subspace of the 
state space ${\cal H}.$ May be such harmonic oscillations are units of mental information processing performed in the QL meta-representation. 

We remark that linear entropy is an approximation of the von Neumann entropy $E(\rho)= \rm{Tr} \rho \ln \rho,$ or in terms of the covariance operator,  
\begin{equation}
\label{VNE}
E(\rho_S)= \frac{1}{\rm{Tr} C_S} \rm{Tr} C_S( \ln C_S - \ln \rm{Tr} C_S I).      
\end{equation}
It is also takes its minimal value for ROs concentrated in one dimensional subspaces of ${\cal H}.$   

\subsection{Quantum-like cognitive observables as quadratic forms of oscillation states} 
\label{QLO}

A cognitive observable $A,$  say a question or a task, is represented by a Hermitian operator, denoted by the same symbol $A.$  The QL model generates the probabilistic predictions for the averages of observables in quantum(-like) states,
\begin{equation}
\label{average}
\langle A\rangle_\rho = \rm{Tr} \rho A.
\end{equation}
QL  operator-observables are determined phenomenologically, typically as matrices matching some probabilistic 
constraints, or with the aid of the algebra of creation and annihilation operators \cite{Bagarello}. The present study connects the QL cognition, specifically decision-making (option selection),   with oscillating neuronal networks in the brain.  We couple the operator-observables with classical oscillations  in a network $S$ of the brain, where $S$ generates an answer to the question $A$ (or performs 
the task $A).$ In this way we demystify the operator representation of observables in QL models. 

Each Hermitian operator $A$ determines the quadratic form $Q_A(z,z)=\langle z|A| z \rangle.$  Consider a random network $S$ with the classical oscillatory state given by a random vector 
$z=z(\omega)$ with the covariance operator $C.$ The quadratic form $Q_A$ determines the random variable $Q_A= Q_A(z(\omega),z(\omega)).$ Then \cite{Beyond,annals} we have 
\begin{equation}
\label{average1}
\frac{1}{\rm{Tr}C} E[Q_A] = \frac{1}{\rm{Tr}C} \rm{Tr} C A=  \rm{Tr} \rho A.   
\end{equation}
In this way a QL observable $A$ is associated with the quadratic form of ROs in $S;$ their averages are coupled by equality (\ref{average1}); the classical average on the left-hand side and the quantum average on the right-hand side.

We note that by the equality (\ref{D}) $\rm{Tr} C$ equals to the dispersion of the random variable
$z=z(\omega)$ (for the variables with zero mean value). Thus, the quantum(-like) average can be represented as the classical average
\begin{equation}
\label{average2}
\langle A\rangle_\rho= \rm{Tr} \rho A =\frac{1}{E[||z||^2]} E[Q_A(z(\omega),z(\omega)].    
\end{equation}

We theorize on the biological meaning of operator observables in QLM.  

\medskip

{\bf Conjecture 3.} {\it The observables (questions, tasks, traits) considered in QLM of cognitive aspects of  decision-making are
coupled to neurophysiological processes in the brain via the correspondence, quadratic form$\to$Hermitian operator.
From this viewpoint, the quantum formalism is the machinery for the linear algebraic representation of the averages 
of the quadratic forms of ROs in neuronal networks.}    

\section{Binding problem  in the light of entanglement}
\label{binding}

The binding problem is one of the fundamental challenges in cognitive science: how are objects, backgrounds, and symbolic or emotional features integrated into a unified experience? It concerns the brain's overall encoding mechanisms for combining decisions, actions, and perception. At present, no universally accepted model exists for binding.

In this section, we move toward resolving the binding problem within the framework of QL oscillatory cognition. In QL terms, binding is simply the entanglement of states. This notion of entanglement necessitates the consideration of compound systems, which, in our framework, correspond to neuronal networks in the brain. Our goal is not merely to describe binding as entanglement formally but to outline how it is generated by these networks. To achieve this, we extend beyond quantum theory by incorporating PCSFT, specifically its classical random field realization of the quantum states of compound systems \cite{Beyond,annals}. This aspect involves complex mathematical formulations, which we present schematically, focusing on its relevance to QL oscillatory cognition (see section \ref{entanglement} for more mathematical details).

{\it Can the QL model of oscillatory cognition provide a solution to the binding problem?}

Yes — provided that the statements in Conjectures 1 and 2 accurately reflect biological reality. According to Conjecture 1, at a meta-level, the brain operates using QL states, functioning as a quantum information processor that operates with qubits rather than classical bits. Conjecture 2 proposes that these QL states correspond to covariance matrices (up to normalization) generated by oscillatory networks. Such representations are integral and nonlocal, meaning that the neuronal circuits contributing to these correlations may be distributed across different brain regions. 

{\it In the QL model of oscillatory cognition, binding occurs within this quantum(-like) information processing.}

Consider the textbook example: creation of a cup image; say  via integration of the  spacial shape and color. There are two networks 
$S_1\equiv S_{\rm{shape}}$ and $S_2\equiv S_{\rm{color}}.$ They generate ROs described mathematically as random vectors 
$z_{1} = (z_{1;j}, j=1,..., N_1)$ and   $z_{2} = (z_{2;j}, j=1,..., N_2).$ These random vectors are valued in Hilbert spaces 
${\cal H}_k, k=1,2.$  In the QL representation the geometric structure and color are described by QL states $\rho_{k} = C_{k}/\rm{Tr} C_{k}.k=1,2,$  where $C_{k}=(E [z_{k;j} \bar{z}_{k;i}]).$  

How can the brain bind these meta-level (quantum information) images given by QL states $\rho_{1}$ and $\rho_{2}?$

The brain binds the shape and color via establishing the correlations between neuronal circuits in $S_1$ and $S_2.$  It produces the integral image of a cup as the quantum state corresponding to correlations in the compound neuronal network $S=(S_1,S_2).$ 
However, this process cannot be described as binding of QL states $\rho_{1}$ and $\rho_{2}.$ In quantum theory the state of a compound system cannot be reconstructed from the states of subsystems. The state of $S,$ given by density matrix
$\rho,$  encapsulates not only the internal correlations within each subsystem but also inter-system correlations that do not exist in the individual subsystem states.

ROs in $S$ generate the covariance matrix $C\equiv C_S.$ This is the covariance matrix of the random vector
$z=(z_{1}, z_{2})$ valued in {\it Cartesian product} ${\cal H}_1 \times {\cal H}_2.$  It has the block structure 
$C= \begin{vmatrix}
C_{11} & C_{12}\\
C_{12}^\star & C_{22}\\
\end{vmatrix},$ where  $C_{kk}=(E[z_{k;i} \bar{z}_{k;j}])$ represents correlations within the subsystem $S_k$ and 
$C_{12}=(E[z_{1;i} \bar{z}_{2;j}])$  represents inter-systems correlations.    This covariance 
matrix has dimension $(N_1+N_2) \times (N_1+N_2).$ But the density matrix $\rho \equiv \rho_S$ of a compound quantum system 
has dimension $N_1 N_2 \times N_1 N_2.$ So, $\rho$  and $C$ cannot be coupled just via scaling   
as in the case of a single network. The main problem is that classical ROs of two systems are described 
in Cartesian product and quantum states in tensor product. PCSFT establishes special connection between $C$ and $\rho$ \cite{Beyond,annals}, see section \ref{entanglement}, where this scheme is presented for pure quantum states. This scheme is based on the operator representation of the state $|\psi\rangle$ of a compound system and identification of this operator with the covariance operator of inter-systems correlations.   

In our example,  the density matrix $\rho_\psi$ represents the image of the cup. Its geometric form and color can be extracted from 
this image and mathematically described as partial traces.  May be the use of the term  ``binding problem'' is ambiguous. At the level of images this is not binding, but extraction of special features of an integral image, in our case the latter is a cup and its special features are the form and color. At the level of neural oscillations ``binding'' has the meaning of establishing correlations between 
ROs in the networks $S_1$ and $S_2.$    

How does the brain establish such correlations? The general answer is ``via electro-chemical signaling''. But propagation of signals via the network of axons and dendrites can be too slow, for distantly located networs $S_1$ and $S_2.$ How does the brain create the long-distance, so to say ``nonlocal'', correlations? We do not know yet... . But mathematics might help us (as sometimes happens in natural science). 

As was shown in \cite{annals} (see section \ref{entanglement}), it is impossible to generate entangled states of $S$ just via correlating  classical ROs in $S_1$ and $S_2.$ The ``covariance   matrix''  behind the entangled state $|\psi\rangle$ is not be positively defined. To make it positively defined, the random vector
$z=(z_{1}, z_{2})$ should be coupled to a background  random field $\Phi.$ This field amplifies the correlations between the random vectors 
$z_{1}, z_{2}.$ From the quantum foundation viewpoint, the quantum correlations are stronger than classical ones. The degree of violation of the Bell inequality, exceeding 2, is coupled to superiority of quantum correlations over classical ones. However, as is shown by PCSFT, quantum correlations can be generated with classical random variables, but through coupling them to the common background field. The presence of such field synchronizes correlations between neurons. Such synchronization induces positively defined matrices of correlations beyond entangled  QL states.\footnote{ 
In quantum physics the background field can be identified with the ``zero-point field'' - the electromagnetic field of vacuum fluctuations. In the brain science the most natural candidates for the background electromagnetic fields are the basic brain rhythms, 
which are classified into alpha rhythm (8–13 Hz), beta (13–35 Hz) and gamma waves (35 Hz and higher). In QL  oscillatory cognition  these rhythms contribute to generation of the entangled QL states. From neurons' location perspective, these states are nonlocal. But this ``nonlocality'' has nothing to do with so widely discussed quantum nonlocality - spooky action at a distance. In PCSFT, ``nonlocality'' is classical, it is generated by the zero point field or in our QL model of socillatory cognition  by brain's rhythms.}

In section \ref{cognitive2} we invented measures of oscillatory cognition and their quantum theoretical counterparts. This approach can be extended to compound neuronal networks. {\it Measures of entanglement can be associated with measures of binding} (section \ref{entanglement}).

We theorize that the binding problem can be solved within the QL model of oscillatory cognition:  

\medskip

{\bf Conjecture 4.} {\it In the QL representation the integral features of an object are represented as entangled states generated 
by oscillatory neuronal networks that are inter-correlated with each other and with the basic background rhythms in teh brain.}

\medskip

We stress that due to Conjecture 1 the brain is able to create the quantum information representations of its states, particularly  entangled states.

\section{From classical to QL  dynamics}
\label{DS}

We stress once again that the procedure of the quantum theoretical representation of the classical Hamiltonian systems should be sharply distinguished from the quantization procedure. 

\subsection{Networks of harmonic oscillators}
\label{DS1}

Now we are interested in the QL lifting of the neuronal dynamics in $S.$ 
Consider a system of interacting harmonic oscillators as the  Hamiltonian system with quadratic Hamiltonian function 
\begin{equation}
\label{H1}
H(q,p)=1/2 [(Rp,p) +2 (Tp,q) + (Rq,q)],
\end{equation}
where $R$ is a symmetric operator, $R^\star= R$  and $T^\star=- T.$ The operator 
\begin{equation}
\label{H1d}
H = \begin{vmatrix}
R&T\\
-T&R\\
\end{vmatrix},
\end{equation}
commutes with the symplectic operator 
\begin{equation}
\label{H1f}
J = \begin{vmatrix}
0&I\\
-I&0\\
\end{vmatrix},
\end{equation}
determining the complex structure on on the phase space $Q\times P.$ We note that any linear operator in ${\cal H}$ commuting with $J$ has the block structure (\ref{H1d}). 

For any Hamiltonian function $H(q,p),$  the system of the Hamiltonian equations has the form
\begin{equation}
\label{H2}
\dot q= \frac{\partial H}{\partial p}(q,p), \; \dot p=- \frac{\partial H}{\partial q}(q,p). 
\end{equation}
The Hamiltonian function (\ref{H1}) the system (\ref{H2}) has the form
\begin{equation}
\label{H3}
\dot q= Rp -Tq, \; \dot p=- (Rq +Tp). 
\end{equation}

This is the system of harmonic oscillators; to see this we write the Hamiltonian system (\ref{H3}) as the second order differential equation for $q=q(t),$ 
\begin{equation}
\label{H3o}
\ddot q + C \dot q + K q=0, 
\end{equation}
where $C=(RTR^{-1} + T), K= (R^2 + RTR^{-1}T).$ If the operators commute. we get $C=2T, K= (R^2+T^2).$  For example, consider two coupled harmonic oscillators with $T=0$ and  $R^2= 
= \begin{vmatrix}
(k+k_{12}) & -k_{12}\\
-k_{12} & (k+k_{12})\\
\end{vmatrix},$ 
where ``spring  stiffness'' coefficients $k, k_{12} >0.$  The system (\ref{H3o}) has the form
\begin{equation}
\label{H3p}
\ddot q_1 =  - k q_1 + k_{12} (q_2-q_1), \; 
\ddot q_2 = - k q_2 + k_{12} (q_1- q_2) . 
\end{equation}
The energy of the first oscillators flow to the second and vice verse.

We can also consider the {\it Wilberforce pendulum} that illustrates so well the electromagnetic oscillator with coupling between the electric and magnetic fields (section \ref{appendix A}). Let $T=0$   and let $R^2= 
= \begin{vmatrix}
k & \epsilon\\
\epsilon & \delta\\
\end{vmatrix},$ where $k\delta - \epsilon^2 > 0.$ In these notation,  the system (\ref{H3o}) has the form
\begin{equation}
\label{H3q}
\ddot q_2 +k q_1 + \epsilon q_2=0, \; \ddot q_2  + \delta q_2 + \epsilon  q_1=0, 
\end{equation}
where $\epsilon$ is the coupling constant between the vertical and rotation oscillators. The energy of the vertical oscillator flows to the rotation  oscillator and vice verse.  
  

Our basic observation \cite{Beyond} is that the classical Hamiltonian system (\ref{H3}) on the phase space $X= Q\times P,$  can be rewritten as the Schr\"odinger equation in the complex space ${\cal H}= Q\oplus iP.$
\begin{equation}
\label{H4}
i \dot z_t= H z_t 
\end{equation}
(see \cite{Beyond}; cf. with the complex Riemann-Silberstein representation of the Maxwell equations, see section \ref{appendix A}). 
Then $z_t=U_t z_0,$ where $U_t= e^{-itH} z_0.$  

For illustration, we restrict randomness within the network $S$ to randomness of initial conditions, i.e., $z_0=z_0(\omega);$
set 
\begin{equation}
\label{H5}
C_0=(c_{0;km}= E[z_{0;k} \bar{z}_{0;m}]). 
\end{equation}
Then 
\begin{equation}
\label{H5a}
C_t=U_t^\star C_0 U_t . 
\end{equation}
We point out that, for such dynamics of the covariance operator, its trace is preserving,
$\rm{Tr} \; C_t=\rm{Tr} \; C_0= \rm{const}.$ The corresponding density operator evolves as 
\begin{equation}
\label{H5b}
\rho_t=U_t^\star \rho_0 U_t .  
\end{equation}
Hence, the density operator $\rho_t$ satisfies the von Neumann equation:
\begin{equation}
\label{H4a}
i \dot \rho_t = [H,\rho_t]  
\end{equation}

Consider now deterministic dynamics of coupled harmonic oscillators; in phase space it generates the trajectory $x_t=(q_{jt}, p_{jt}))$ starting with non-random initial condition   $x_0=(q_{j0}, p_{j0}) \in \mathbb{R}^{2N}.$  The corresponding dynamics in $N$ dimensional complex space has the trajectory $z_t=(z_{jt})$ starting with the initial condition 
$z_0=(z_{j0}) \in \mathbb{C}^N,$ where $z_{j0}= q_{j0} +i p_{j0}, k=1,...,N.$ The covariance matrix of the non-random initial condition has the form $C_{z_0}=|z_0\rangle  \langle z_0|.$  Thus deterministic dynamics of coupled harmonic oscillators is mapped to the quantum pure state dynamics. The same happens for random initial conditions concentrated in the one dimensional subspace ${\cal H}_{z_0}.$

This representation can be also constructed for Hamiltonian function $H(q,p)=\frac{1}{2} [(R_1 p,p) +(R_2q,q)].$ Here the kinetic and potential energies are described by two different quadratic forms.  
The Hamiltonian  dynamics 
\begin{equation}
\label{H3g}
\dot q= R_1p, \; \dot p=- R_2q
\end{equation}
 can be represented as the Schr\"odinger dynamics - at least in the case of commuting matrices $R_1,R_2.$ We introduce the complex variable 
by setting 
\begin{equation}
\label{H3g1}
z_t= \sqrt{R_2} q_t +  i \sqrt{R_1} p_t.
\end{equation}
 Then, under the assumption $[R_1,R_2]=0, \; z_t$ satisfies the Schr\"odinger equation (\ref{H4}) with the Hamiltonian 
$H = \sqrt{R_1R_2}= \sqrt{R_2R_1}.$  

The approach presented in this section can be generalized to time dependent Hamiltonian functions
$H(t;q,p)=1/2 [(R(t)p,p) +2 (T(t)p,q) + (R(t)q,q)],$ where for each $t$ the operator (\ref{H1d}), $H=H(t),$ commute with symplectic 
operator $J.$ Such Hamiltonian dynamics lead to the Schr\"odinger dynamics 
with time dependent Hamiltonians. In particular, we obtain the system of coupled harmonic oscillators as (\ref{H3p}) with 
time dependent ``spring stiffness'' coefficients $k=k(t), k_{ij}=  k_{ij}(t).$  We remark that such systems are explored 
in oscillatory based QIC (section \ref {QICS}), e.g. equations (6), (7) in article \cite{Goto} devoted to solution of the Ising problem - to find a spin configuration minimizing the Ising energy).   

We conclude that the dynamics of any network of coupled harmonic oscillators can be represented in the QL way. Linear 
oscillators are characteristic for an isolated network of nodes. Interactions with other networks generate more complex oscillations in nodes (section \ref{Gnetwork}).    

\section{Oscillatory networks in quantum inspired and cognitive computations}
\label{QICS}

The authors of article \cite{Wolf} rightly pointed to the analogy between oscillatory information processing in the brain and QIC. What is QIC?  We cite review \cite{QIC} on QIC:

\begin{footnotesize} ``The term ``quantum-inspired'' was introduced by Moore
and Narayanan in the context of computing for the first
time in 1995. The term was used to differentiate between
two types of computational methods: ``pure'' quantum computation and quantum-inspired computation. The former
is firmly rooted in quantum mechanical concepts, such as
standing waves, interference, and coherence, and can only
be executed on a quantum computer. On the other hand,
``quantum-inspired'' computing refers to practical methods
that have been derived from these concepts. These methods do not require a quantum computer, but rather utilize
classical computers or algorithms to simulate quantum effects and achieve computational advantages.'' 
\end{footnotesize}   

There are both similarities and differences between QL cognition and QIC. Neither approach refers to genuine quantum physical processes.
The main distinction between QL cognition and QIC is that QL directly employs the theoretical formalism of quantum theory, while QIC uses methods that share some similarities with quantum theory, specifically the aspects applied in quantum computing and simulations.

QL cognition utilizes quantum formalism to describe fundamental effects in cognition and decision-making. By constructing a QL model based on complex Hilbert space and quantum measurement theory, one can leverage the quantum phenomena predicted by this formalism, such as order and interference effects in decision-making. The ``only challenge'' is to justify the use of this formalism. We hope that this paper, along with the recent work \cite{GPT}, provides such justification by demonstrating its connection to the functioning of oscillatory networks in the brain (see also \cite{DBS,DB,TC,BUSN,Scholes1,Scholes2}).

In contrast to QL, QIC does not directly explore the quantum formalism - quantumness is merely a useful metaphor. As a result, QIC sidesteps the main challenge faced by QL, which is justifying the use of quantum theory. However, the trade-off is that QIC must provide a non-quantum explanation for its similarities to quantum computation (or quantum simulators) and, above all, justify the claim of its superior computational power compared to traditional classical digital and simulator computations. At its current stage, QIC cannot fully provide such justification; it mainly relies on numerical simulations, which are treated as evidence for the claimed superiority \cite{QIC, Goto0, Goto}.

We now turn to a specific area of QIC: computational algorithms based on networks of classical oscillators, e.g., \cite{Goto0, Goto}. In this framework, classical oscillators mimic qubits by being in a "superposition-like" state, represented as being "here and there." We have developed a QL representation for oscillatory networks. This representation can be used not only to justify QL cognition but also to support the claim of superiority in QIC.

\section{Quantum-like AI based on oscillatory networks and quantum inspired computation}
\label{AI}

Humanoid AI should be based on QL model. One of the lessons of QL cognitive studies is that some distinguishing features of human cognition, as e.g. the order, conjunction, disjunction, and response replicability effects 
as well contextuality,   match well with quantum theory. Hence humanoid AI systems should process information by respecting the 
laws of quantum theory. Of course, this can be done with genuine quantum physical computers. But, for the moment, such devices are far from being useful for the real technological purposes, e.g. creation of AI systems.\footnote{The level of noisy outputs of modern quantum computer is too high. To down the noise to an acceptable level, each logical qubit should be implemented on the basis of approximately 1000 physical qubits. Thus, for quantum computer with 1000 of such logical qubits, one needs 1 million of physical qubits ...} On the other hand, QL representation of functioning of oscillatory network is theoretically identical to the one explored in quantum physics. This opens the door to QLAI based on classical analog devices with implemented oscillatory networks. For example, such devices can serve as the physical base for humanoid robots. Oscillatory networks can also be implemented on classical digital computers and serve as the basis for humanoid avatars.   

The algorithms designed in QIC and based on oscillatory networks can  be employed for realization of QLAI. In this case AI designer is interested not only in the computational power of such algorithms, but also (and in some cases mainly) in realization of QL features of AI system. Thus, QLAI  is a new perspective area of applications of QIC.

\section{Concluding remarks} 

We constructed a bridge between QL features of cognition observed and explored in decision making, behavioral and social science, economics and finance, and information processing in oscillatory neuronal networks in the brain. Consideration of these networks as the biological basis of the QL representation was stimulated by the study  \cite{Wolf} in which 
neuronal oscillatory circuits were tested as computational devices similar to ones explored in QIC.
The present paper is closely related to article \cite{GPT} in that QL cognition was supported by representations of neuronal activity within generalized probability theory (operational quantum measurement theory). In \cite{GPT} the QL representation  
is over the field of real numbers and now we designed the complex Hilbert space picture.\footnote{The complex picture is especially important for 
electric oscillators generating the electromagnetic field $\Phi=(E, B),$ with electric and magnetic components. The complex description of this field $Z= E+ i B$ is the well known Riemann-Silberstein expression for the electromagnetic field). On the other hand, the real models created in \cite{GPT} are not less interesting. For example, EEG's output is the real-valued electric potential $V.$ In \cite{GPT} we considered  the real-valued
action potential and frequencies of oscillations as the outputs of the nodes of neuronal networks.
By turning to EEG and real valued QL representations, we can say that the routine EEG measurements can be used for testing 
the basics of our theory, namely, the coupling between classical ROs in btain's networks and quantum formulas for averages, starting with calculation of covariance matrices for EEG signals. But this is the project for further studies. }   
     
As was noted, the mathematical scheme works not only for biological, but even for arbitrary physical oscillators and it can be used for foundational support of QIC. But the latter is the separate project - to connect QIC superiority with QL representation constructed in this article. Finally, we point out (see section \ref{AI}) that coupling of oscillatory calculations in the brain 
\cite{Wolf} with QL cognition and QIC may lead to a new sort of AI creatures, e.g., robots and avatars, namely QLAI systems.        

\section*{Acknowledgments} 

A.K. was partially supported by  EU-grant CA21169 (DYNALIFE) and visiting professor fellowship at the group of prof. Iriyama at  Tokyo University of Science (April 2024). A.I. was supported by Grant-in-Aid for Transformative Research Areas, N24H02200, Japan Society for Promotion of Science.
                
\section{Supplementary material}
\label{supp}

\subsection{Compound systems: connection between covariance operator  and quantum state}
\label{entanglement}

Here we follow \cite{Beyond,annals}. To simply consideration, we consider the case of a pure quantum state of a compound system $S=(S_1,S_2),\;   |\psi \rangle  \in {\cal H}_1 \otimes {\cal H}_2, ||\psi||=1,$ that is the corresponding density operator $\rho_\psi=|\psi \rangle \langle \psi|$ is one dimensional projection. We note that in this case $\rho_1$  and $\rho_2$ are mixed states. In PCSFT the following identification is employed ${\cal H}_1 \otimes {\cal H}_2 \cong 
{\cal L} ({\cal H}_2, {\cal H}_1),$ the latter is the space of linear operators acting from ${\cal H}_2 \to {\cal H}_1.$  Hence each pure state of compound system can be represented as a linear operator $L_\psi.$  We point out that such identification was employed by von Neumann in his mathematical construction of entanglement \cite{VN}. 

The starting point of the PCSFT description of the classical ROs behind a pure state $|\psi\rangle$ is the appeal to the following candidate for the  covariance matrix of the random vector $z=(z_1,z_2),$ 
\begin{equation}
\label{LL}
D_\psi = \begin{vmatrix}
L_\psi L_\psi^\star & \: \; L_\psi\\
\: \;  L_\psi^\star &   L_\psi^\star L_\psi\\
\end{vmatrix}.
\end{equation}
In such construction a pure quantum state $|\psi\rangle$ of a compound system $S=(S_1,S_2)$ represents inter-systems correlations, $D_{12}= 
L_\psi.$ Correlations within subsystems age given by operators
\begin{equation}
\label{LLg}
D_{11} = L_\psi L_\psi^\star = \rm{Tr}_{{\cal H}_2} \rho_\psi=\rho_1,
D_{22} =L_\psi^\star L_\psi = \rm{Tr}_{{\cal H}_1} \rho_\psi = \rho_2.
\end{equation}
Here we need not normalize  $D_{11}$ and $D_{22}$ of ROs in $S_1$ and $S_2,$ because 
$\rm{Tr} D_{11}  =\rm{Tr} D_{22} = ||\psi||^2=1,$ We started with the quantum pure state $\psi$ and wanted to construct 
ROs behind it. We can start with an arbitrary vector $\Psi$ representing inter-network correlations and proceed to 
QL representation of ROs. Then we shall operate with normalization $|\psi \rangle=  \Psi/||\Psi||.$  

We note that, for $h_1 \in {\cal H}_1, h_2 \in {\cal H}_2,$  $E[\langle h_1| z_1\rangle \langle z_2| h_2\rangle]= 
\langle L_\psi h_2| h_1\rangle = \langle \psi| h_1 \otimes h_2\rangle.$ If state $\psi$ is non-entangled (factorizable)
$|\psi\rangle = |\psi_1\rangle \times |\psi_2\rangle,$ then the inter-systems correlations are also factorisable:  
$E[\langle h_1| z_1\rangle \langle z_2| h_2\rangle] = E[\langle h_1| z_1\rangle] E[\langle z_2| h_2\rangle].$ For jointly 
Gaussian circularly symmetric random vectors, this equivalent to their independence. Hence non-factorization  of inter-systems
correlations corresponds to an entangled state $|\psi\rangle.$     

However, as was proved in \cite{annals}, the operator $D_\psi$ is positively defined only for a non-entangled state, $|\psi\rangle =
|\psi_1\rangle \otimes |\psi_2\rangle.$ To generate entangled states, one should add to 
$z=(z_1,z_2)$ another random vector that can be interpreted as a background random field (section \ref{binding}). The  simplest deformation of $D_\psi$ is given by additional white noise that should be sufficiently strong to lead to positively defined covariance operator $C_\psi= D_\psi + \epsilon I.$ This is just mathematics, but it has an important foundational consequence (section \ref{binding}).

One of the important measures of entanglement is von Neumann entropy of the density operators of the subsystems of a compound system,  
$E_k= \rm{Tr} \rho_k \ln \rho_k.$ For simplicity, we again proceed with a pure state $|\psi\rangle.$ Here, e.g.,  
$E_1= \rm{Tr} L_\psi L_\psi^\star \ln L_\psi L_\psi^\star.$  For an entangled state, $E(\rho_1) >0.$ 

Now we should recall that the operator $D_\psi$ is not positively defined and the covariance operator $C_\psi$ should include the contribution of the background (electromagnetic) oscillations. Mathematically  simplest model of such background leads to 
$C_\psi= D_\psi + \epsilon I,$ or $\rho_i = C_{ii} - \epsilon I.$ Hence, in terms of classical ROs correlations,  
$E_i=  \rm{Tr}  (C_{ii} - \epsilon I) \ln (C_{ii} - \epsilon I).$ So, if $E_i>0,$ then 
the state $|\psi\rangle$ is entangled. In our model of binding this means that  the  
images generated by the neuronal networks are bound.  

Linear entropy is often used as approximation of von Neumann entropy, as a measure of entanglement, in terms of the covariance operator
$E_{L,i}= 1- \rm{Tr} (C_{ii} - \epsilon I)^2.$

\subsection{Quantum-like representation for the classical electromagnetic field}
\label{appendix A}
Consider the dynamics  of the classical electromagnetic field $\phi(t,x)=(E(t,x),B(t,x)).$
Consider the functional space $F$ of square integrable vector-valued (in $\mathbb{R}^3)$  functions constrained by the equation 
$\nabla \times f(x)=0.$ This space is constructed as the $L_2$-completion of the space of infinitely differentiable functions with 
compact support constrained by the equations $\nabla \times f(x)=0.$ We set $Q=F$ and $P=F$ and work in the phase space $Q\times P$ of fields 
$\phi(x)=(E(x),B(x))$ that are square integrable and constrained by the system of equations 
\begin{equation}
\label{HRS0}
\nabla \times E(x)=0 , \;  \nabla \times B(x)=0.
\end{equation}
Consider the quadratic Hamilton function defined on this phase space 
\begin{equation}
\label{HRS}
H(E,B)= \frac{c}{2} \int_{\mathbb{R}^3} \Big[ (\nabla \times E(x)|E(x)) + (\nabla \times B(x)|B(x))   \Big] dx.
\end{equation}
where $c$ is the light velocity and $(\cdot|\cdot)$ is the integral scalar product.
The corresponding (infinite dimensional)  system of the Hamiltonian equations has the form
\begin{equation}
\label{HRS1}
\frac{1}{c} \frac{\partial E}{\partial t}(t,x) = \nabla \times B(t,x),
\end{equation}
\begin{equation}
\label{HRS2}
\frac{1}{c} \frac{\partial B}{\partial t}(t,x) = \nabla \times E(t,x).
\end{equation}
with the initial condition $\phi(0,x)=(E_0(x),B_0(x)).$ 
This is the system of Maxwell equations.  Now consider the complex (Riemann-Silberstein) representation of the electromagnetic field, 
$Z(t,x)= E(t,x) + i B(t,x).$  Then $Z(t,x)$  satisfies the Schr\"odinger equation
\begin{equation}
\label{HRS1a}
i \frac{\partial Z}{\partial t}(t,x)  = H Z(t,x),
\end{equation}
where  ``Hamiltonian'' has the form $H = c \nabla \times.$

ROs in initial conditions, $Z_0=Z_0(x; \omega)=E_0(x;\omega) + i B_0(x;\omega)$ lead to stochastic process, $Z(t,x; \omega).$ Its covariance operator $C_t$ acting in  $L_2(\mathbb{R}^3)$ is defined as
$$
\langle C_t \phi_1|\phi_2\rangle= 
\int_{\mathbb{R}^6} E[Z(t,x) \bar{Z}(t,y)] \bar{\phi}_1(x) \phi_2(y) dx dy,
$$
so this is the integral operator with the kernel 
$K(t,x, y)= E[Z(t,x) \bar{Z}(t,y)].$           
This is Hermitian positively defined operator with trace
$\rm{Tr} C_t=  \int_{\mathbb{R}^3} K(t,x,x) dx =
\rm{Tr} C_0= \int_{\mathbb{R}^3} K(0, x,x) dx.$ The corresponding density operator $\rho_t$ has the kernel 
$\frac{1}{\rm{Tr} C_0} K(t,x, y).$ It satisfies the von 
Neumann equation.

\subsection{Quantum-like representation for damped harmonic oscillator}
\label{appendix B}

Consider the damped harmonic oscillator in one dimensional case 
\begin{equation}
\label{DO}
m \ddot q+ \beta \dot q +kq=0, \; \mbox{or} \; 
\ddot q+ \alpha \dot q + \omega_0^2 q=0, 
\end{equation}
where $\alpha= \beta/m,  \omega_0^2= k/m.$
This dynamics can be written in the form of Hamiltonian dynamics with the Hamilton function, 
\begin{equation}
\label{DO1}
H(t; q,p)= \frac{p^2}{2m} e^{-\alpha t} + \frac{kq^2}{2} e^{\alpha t},
\end{equation}
where the momentum variable $p= m \dot q e^{\alpha t}.$  The system of Hamiltonian equations has the form  
\begin{equation}
\label{DO2}
\dot q_t= e^{-\alpha t} p_t/m , \; \dot p_t=-  k e^{\alpha t}  q_t. 
\end{equation}
Really, $\ddot q_t= - \alpha e^{-\alpha t} p_t/m +  e^{-\alpha t} \dot p_t/m = - \alpha \dot q_t - (k/m) q_t$ or 
$\ddot q_t + \alpha \dot q_t + (k/m) q_t =0,$ and we recognize the second of equations (\ref{DO}). 

Set  $z_t= q_t + ip_t.$  However, as can be expected for the damped oscillator, this complex valued function doesn't satisfy to any 
Schr\"odinger equation of the form $i \dot z_t= H(t) z_t,$ where $H(t)$ is the real valued function. Nevertheless, we can perform 
transformation of the phase-space coordinates leading to the  Schr\"odinger equation. 

In (\ref{DO2}) set $a=1/m,b=k.$ Consider the following complex linear time dependent transformation of the phase plane, 
$z_t= f_t q_t + ig_t p_t,$ where $f_t= e^{\alpha t/2} f_0,  g_t= e^{-\alpha t/2} g_0,$ where $f_0, g_0$ are complex constants. We want to prove that there exist constants and the real non-negative number $\Gamma$ such that $z_t$ satisfies the one dimensional Schr\"odinger equation:
\begin{equation}
\label{DO3}
\dot z_t= \Gamma z_t. 
\end{equation}
 Straightforward calculation leads to the system of two linear equation for constants $f_0, g_0,$
 \begin{equation}
\label{DO4}
i\alpha f_0 +2b g_0= 2 \Gamma f_0, \; - i\alpha g_0 +2a f_0= 2 \Gamma g_0, 
\end{equation}
or 
\begin{equation}
\label{DO5}
(i\alpha - 2\Gamma) f_0 +2b g_0= 0, \;  
2a f_0 - (i\alpha + 2 \Gamma) g_0 = 0, 
\end{equation}
The condition existence of its non-zero solution determines the ``Hamiltonian'' $\Gamma.$ We have condition 
$\Gamma^2=  \alpha^2/4 + ab$ and $ \Gamma^2=  \sqrt{\alpha^2/4 + ab}.$ Similar calculations can be performed in 
the multi-dimensional case.  
 

\subsection{From linearly evolving covariance operator to nonlinearly evolving density operator}
\label{Gnetwork}

We consider networks of neuronal circuits with linearly evolving covariance operators. This is a classical counterpart of quantum theory of open systems. Generally such networks are not isolated and they interact with other circuits networks in the brain and body. In the quantum formalism such covariance dynamics corresponds to generally nonlinear master equation,  (\ref{L2}) 
(see \cite{KHRBNL}, cf. \cite{Ottinger1,Ottinger2,Tsekov}). In the particular case of linear dynamics (\ref{L3}) we obtain, for example, the GKSL equation. It is important to remind, that the covariance dynamics is the image of a class of classical stochastic processes valued in a complex Hilbert space, 
$C_t= E [z_t \bar{z}_t].$ Here $z_t$ isn't uniquely determined. We repeat once again that the QL representation is just a shadow of the ``ontic classical model'', its epistemic image.    

The only difference between complex covariance operators and density operators is that the latter are normed by trace one. 
The normalization procedure (\ref{lll1}) transfers covariance operators to density operators. We want to establish connection between two dynamics, for a covariance operator and the corresponding density operator, between $C_t$ and $\rho_t= C_t/\rm{Tr} C_t.$ This correspondence was established in article \cite{KHRBNL} and we briefly present some essential moments. 

Denote by ${\cal L}$ a superoperator that is the generator a semigroup of maps $V_t = e^{t{\cal L}}$ such that 
\begin{equation}
\label{L}
 V_t : {\cal C} \to {\cal C}.
\end{equation}
The general form of  ${\cal L}$ isn't known. But there are numerous known
examples of such generators;  for example, the generator of the GKSL equation. 
We set $C_t = V_t C_0,$ where $C_0$  is a positive Hermitian operator. Then $C_t$ is one parametric group of positive Hermitian 
operators determining the solutions of the following Cauchy problem
\begin{equation}
\label{L1}
 \dot C_t= {\cal L} C_t, \; C_{t=0}=C_0. 
\end{equation}
As was shown in \cite{KHRBNL}, the corresponding density operator satisfies to the following equation
\begin{equation}
\label{L2}
 \dot \rho= {\cal L} \rho - \rho \rm{Tr} {\cal L} \rho  , \; \rho_{t=0}=\rho_0. 
\end{equation}
Thus the ``master equation'' corresponding to general covariance operator dynamics (\ref{L1}) is nonlinear. If the covariance operator dynamics is trace preserving,  then 
$$
0= \frac{d}{dt} \rm{Tr} \rho_t=  \rm{Tr} \frac{d}{dt} \rho_t= \rm{Tr} {\cal L} \rho_t,
$$  
and dynamics (\ref{L1}) becomes linear: 
\begin{equation}
\label{L3}
\dot \rho_t= {\cal L} \rho_t, \; \rho_{t=0}=\rho_0. 
\end{equation}
Suppose that ${\cal L}$ is the generator of the GKSL dynamics that is 
\begin{equation}
\label{GKSLL}
 {\cal L} \rho= - i [H,  \rho] +\sum_j \gamma_j \Big(  A_j \rho A_j^\star - \frac{1}{2} \{A_j^\star A_j, \rho\}  \Big),
\end{equation}
where $H$ is a Hermitian operator and $A_j$ are arbitrary operators describing the interaction of a system $S$ with its physicochemical  environment and 
$\gamma_j\geq 0$ are coupling constants. Then trace is preserved and after normalization by the trace the solution $C_t$ of equation (\ref{L1}) satisfies the GKSL equation (\ref{L3}) describing the evolution of the quantum state. In theory of open quantum systems 
the GKSL equation  describes the state evolution for a quantum system $S.$ In the model under consideration $S$ is a 
classical system. ``Quantumness'' is the property of the representation. At the same time our QL approach doesn't deny 
the possible contribution of the genuine quantum physical processes in the brain and body, i.e.,   the chain of representations,
\medskip
genuine quantum$\to$classical$\to$quantum-like.   
\medskip

If a random network $S$ generates non-trace preserving dynamics of its covariance operator then the corresponding  QL dynamics (\ref{L2})  is non-linear. We point out that nonlinear master equations are widely used in theory of open quantum systems \cite{Tsekov}, e.g., the nonlinear thermodynamic quantum master equation \cite{Ottinger1}. Moreover, numerical solutions of such nonlinear quantum master equations can be found through correlations of classical stochastic processes valued in a complex Hilbert space \cite{Ottinger2}, precisely as in our approach.



\begin{thebibliography}{199}


\bibitem{Aequbitts0} Aerts, D., Broekaert, J., Gabora, L., \& Sozzo, S. (2013). Quantum structure and human thought. Behavioral and Brain Sciences, 36(3), 274-276.

\bibitem{Aerts} Aerts, D.; Gabora, L.; Sozzo, S. Concepts and their dynamics: A quantum-theoretic modeling of human thought. Top. Cogn. Sci. 2013, 5, 737-772

\bibitem{ASR} Alodjants, A. P., Tsarev, D. V., Avdyushina, A. E., Khrennikov, A. Y., \& Boukhanovsky, A. V. (2024). Quantum-inspired modeling of distributed intelligence systems with artificial intelligent agents self-organization. Sc. Rep. 14(1), 15438

\bibitem{Scholes2} Amati, G., \& Scholes, G. D. (2024). Quantum information with quantum-like bits. arXiv preprint arXiv:2408.06485.

\bibitem{QBIOP} Arndt, M., Juffmann, T., and Vedral, V.: Quantum physics meets biology. HFSP J \textbf{3}  386--400 {2009}.

\bibitem{Asano0} Asano, M., Basieva, I., Khrennikov, A., Ohya, M., Tanaka, Y., \& Yamato, I. (2012). Quantum-like model for the adaptive dynamics of the genetic regulation of E-coli's metabolism of glucose/lactose. Systems and synthetic biology, 6, 1-7.

\bibitem{Asano} Asano M, Basieva I, Khrennikov A, Ohya M, Tanaka Y, Yamato I. (2013) A model of epigenetic evolution based on theory of open quantum systems. Systems and Synthetic Biology, 7(4):161-73.

\bibitem{QIB} M. Asano, I. Basieva, A. Khrennikov, M. Ohya, Y. Tanaka, I. Yamato Quantum Information Biology: from information interpretation of quantum mechanics to applications in molecular biology and cognitive psychology. {\it Found. Phys.} 
45, N 10, 1362-1378 (2015).

\bibitem{QL3} Asano, M., Khrennikov, A., Ohya, M., Tanaka, Y. and Yamato, I.: Quantum Adaptivity in Biology: from Genetics to Cognition. Springer, Heidelberg-Berlin-New York (2015).

\bibitem{AsanoD} Asano, M., Basieva, I., Khrennikov, A., \& Yamato, I. (2017). A model of differentiation in quantum bioinformatics. Progress in Biophysics and Molecular Biology, 130, 88-98.


\bibitem{Bagarello} Bagarello, F. Quantum Concepts in the Social, Ecological and Biological Sciences; Cambridge University Press: Cambridge, UK, 2019.

\bibitem{Bagarello1} Bagarello, F., Gargano, F., \& Oliveri, F. (2023). Quantum tools for macroscopic systems. Cham, Switzerland: Springer.

\bibitem{Bagarello2} Bagarello, F., Gargano, F., Gorgone, M., \& Oliveri, F. (2023). Spreading of information on a network: a quantum view. Entropy, 25(10), 1438.
 
\bibitem{Behti} Basieva, I., Cervantes, V. H., Dzhafarov, E. N., \& Khrennikov, A. (2019). True contextuality beats direct influences in human decision making. Journal of Experimental Psychology: General, 148(11), 1925.

\bibitem{BioBas} I. Basieva, A. Khrennikov, M. Ozawa, Quantum-like modeling in biology with open quantum systems and instruments,
Biosystems, 201, 2021, 104328.

\bibitem{Bruza}  Bruza, P. D., Kitto, K., Ramm, B. J., and Sitbon, L. (2015). A probabilistic framework for analysing the compositionality of conceptual combinations. J. Math. Psych., 67, 26-38.

\bibitem{Bruza1} Bruza, P. D., Wang, Z., \& Busemeyer, J. R. (2015). Quantum cognition: a new theoretical approach to psychology. Trends in cognitive sciences, 19(7), 383-393.

\bibitem{Bruza2} Bruza, P. D., Fell, L., Hoyte, P., Dehdashti, S., Obeid, A., Gibson, A., \& Moreira, C. (2023). Contextuality and context-sensitivity in probabilistic models of cognition. Cognitive Psychology, 140, 101529.

\bibitem{B0} Busemeyer, J. R., Wang, Z., \& Townsend, J. T. (2006). Quantum dynamics of human decision-making. J. Math. Psych. 50(3), 220-241.

\bibitem{Busemeyer} Busemeyer, J.R.; Bruza, P.D. Quantum Models of Cognition and Decision; Cambridge University Press: Cambridge, 
UK, 2012; 2nd Edition, 2024.

\bibitem{BUSN} J. R. Busemeyer, P. Fakhari, P. Kvam, Neural implementation of operations used in quantum cognition,
Prog. Biophys. Molecular Biology, 130, Part A, 2017, 53-60.

\bibitem{Davies-Lewis} Davies, E. B. and Lewis, J. T. An operational approach to quantum probability. Commun. Math. Phys. 17, 239–260 (1970).
\bibitem{DV}   Davies, E. B.  {\it Quantum theory of open systems},  (Academic Press,  London, 1976).

\bibitem{DBS} De Barros, J. A., \& Suppes, P. (2009). Quantum mechanics, interference, and the brain. J. Math. Psych, 53(5), 306-313.

\bibitem{DB}  J. A. De Barros, Quantum-like model of behavioral response computation using neural oscillators,
Biosystems, 110,  2012, 171-182.


\bibitem{Wolf}  F. Effenberger, P. Carvalho, I. Dubinin, W. Singer, The functional role of oscillatory dynamics in neocortical circuits: a computational perspective. Proceedings of the National Academy of Sciences, 122(4), e2412830122.


\bibitem{Kvam} Epping, G. P., Kvam, P. D., Pleskac, T. J., \& Busemeyer, J. R. (2023). Open system model of choice and response time. 
J. Choice Mod, 49, 100453.

\bibitem{Pothos23} Epping, G. P., Fisher, E. L., Zeleznikow‐Johnston, A. M., Pothos, E. M., \& Tsuchiya, N. (2023). A quantum geometric framework for modeling color similarity judgments. Cognitive Science, 47(1), e13231.

\bibitem{Gallus} Gallus, C., Pothos, E. M., Ebelt, Z., Skouteris, D., \& Blasiak, P. (2024). Probability Updating in the Classical and the Quantum Model. Quantum Economics and Finance, 1(2), 122-137.


\bibitem{Goto0} H. Goto, K. Tatsumura, and A. R. Dixon, Combinatorial optimization by simulating adiabatic bifurcations in nonlinear Hamiltonian systems.  Sc. Advances, 2019, 5, N 4, 1-8.
\bibitem{Goto}  H. Goto, K. Endo, M. Suzuki, Y. Sakai, T. Kanao, Y. Hamakawa, R. Hidaka, M. Yamasaki, and K. Tatsumura, 
High-performance combinatorial optimization based on classical mechanics. Sc. Advances, 2021,
7, N 6, 1-9.

\bibitem{SingerG1} Gray, C. M., \& Singer, W. (1989). Stimulus-specific neuronal oscillations in orientation columns of cat visual cortex. Proceedings of the National Academy of Sciences of the USA, 86, 1698-1702.
\bibitem{SingerG2} Gray, C. M., K\"onig, P., Engel, A. K., \& W. Singer, W. (1989). Oscillatory responses in cat visual cortex exhibit inter-columnar synchronization which reflects global stimulus properties. Nature, 338, 334-337


\bibitem{Gunji1}  Gunji YP, Shinohara S., Haruna T., Basios V. Inverse Bayesian inference as a key of consciousness featuring a macroscopic quantum logical structure. Biosystems. 2017,  ;152:44-65.
\bibitem{Gunji2} Gunji YP, Sonoda K, Basios V. Quantum cognition based on an ambiguous representation derived from a rough set approximation. Biosystems. 2016 Mar;141:55-66. 

\bibitem{H} Hameroff, S. (1994). Quantum coherence in microtubules. A neural basis for emergent consciousness?  
\emph{J. Cons. Stud.}, 1, 91--118.

\bibitem{Haven} Haven, E.; Khrennikov, A. Quantum Social Science; Cambridge University Press: Cambridge, UK, 2013. 
\bibitem{QLH} Haven, E.,  Khrennikov, A. and Robinson, T. R.: Quantum Methods in Social Science: A First Course. WSP, Singapore  (2017)
\bibitem{handbook}  Haven, E. and Khrennikov, A.: The Palgrave handbook of quantum models in social science. Macmillan Publishers Ltd:  London (2017) 

\bibitem{Holik} Holik, F., Massri, C., Plastino, A., and Saenz, M. (2021). Generalized probabilities in statistical theories. Quantum Rep. 3(3), 389-416.


\bibitem{QIC} L. Huynh, J. Hong, A. Mian, H. Suzuki, Y. Wu, S. Camtepe, Quantum-inspired machine learning: a survey. 	arXiv:2308.11269 [cs.LG].

\bibitem{Igamberdiev1} Igamberdiev, A. U. (1993). Quantum mechanical properties of biosystems: a framework for complexity, structural stability, and transformations. Biosystems, 31(1), 65-73.

\bibitem{Igamberdiev2} Igamberdiev A. U. (2004),  Quantum computation, non-demolition measurements, and reflective control in living systems. Biosystems 77, 47-56.

\bibitem{Igamberdiev2a} A. U. Igamberdiev, J. E. Brenner, Mathematics in biological reality: The emergence of natural computation in living systems Biosystems 204, 2021, 104395.

\bibitem{Iriki0} Iriki, A., \& Taoka, M. (2012). Triadic (ecological, neural, cognitive) niche construction: a scenario of human brain evolution extrapolating tool use and language from the control of reaching actions. Philosophical Transactions of the Royal Society B: Biological Sciences, 367(1585), 10-23.

\bibitem{Iriki} Iriki, A., Suzuki, H., Tanaka, S., Vieira, R. B., \& Yamazaki, Y. (2021). The sapient paradox and the great journey: Insights from cognitive psychology, neurobiology, and phenomenology. Psychologia, 63(2), 151-173.

\bibitem{Iriki1} Iriki, A., \& Tanaka, S. (2024). Potential of the Path Integral and Quantum Computing for the Study of Humanities: An Underlying Principle of Human Evolution and the Function of Consciousness. Global Perspectives, 5(1).

\bibitem{pMENTAL} Khrennikov, A. (2000). Classical and quantum dynamics on $p$-adic trees of ideas. BioSystems, 56(2-3), 95-120.
\bibitem{Khrennikov2002}  Khrennikov, A. (2002). Brain as a quantum-like computer. BioSystems, 64(1–3), 23–31.
\bibitem{KHC3}  Khrennikov, A.  (2004). On quantum-like probabilistic structure of mental information, 
\emph{Open Systems and Information Dynamics} 11 (3), 267--275.
\bibitem{QL0} Khrennikov, A.: Information Dynamics in Cognitive, Psychological, Social, and Anomalous Phenomena. Ser.:
Fundamental Theories of Physics. Kluwer, Dordrecht (2004)

\bibitem{KHR_IM} Khrennikov A. (2006) Quantum-like brain: ``Interference of minds''. Biosystems,  84(3),225-41

\bibitem{UB_KHR} Khrennikov, A. Ubiquitous Quantum Structure: From Psychology to Finances; Springer: Berlin/Heidelberg, Germany; New York, NY, USA, 2010

\bibitem{Khrennikov2010}  Khrennikov, A. (2010). On the physical basis of theory of ``mental waves''. 
Neuroquantology 8. S71–S80.

\bibitem{Beyond} Khrennikov, A. (2014). Beyond quantum. Singapore: Pan Stanford Publ.

\bibitem{KHRBNL} Khrennikov, A., Basieva, I. (2014) Quantum-State Dynamics as Linear Representation of Classical (Nonlinear) Stochastic Dynamics. J Russ Laser Res 35, 71–78.

\bibitem{PLOS} Khrennikov, A., Basieva, I., Dzhafarov, E. N., \& Busemeyer, J. R. (2014). Quantum models for psychological measurements: an unsolved problem. PloS one, 9(10), e110909.

\bibitem{SR} Khrennikov, A., Basieva, I., Pothos, E. M., and Yamato, I. (2018). Quantum probability in decision making from quantum information representation of neuronal states. Scientific Reports, 8(1), 16225.


\bibitem{Open_KHR} Khrennikov, A. Y. (2023). Open Quantum Systems in Biology, Cognitive and Social Sciences. Springer Nature, Berlin.


\bibitem{Khrennikov} Khrennikov A. Open Systems, Quantum Probability, and Logic for Quantum-like Modeling in Biology, Cognition, and Decision-Making. Entropy. 2023; 25(6):886. 

\bibitem{annals} Khrennikov, A. (2024). Characterization of entanglement via non‐existence of a subquantum random field. Annalen der Physik, 536(9), 2400035.

\bibitem{Iryama} Khrennikov, A., Iriyama, S., Basieva, I., \& Sato, K. (2024). Quantum-like environment adaptive model for creation of phenotype. BioSystems, 242, 105261.


\bibitem{GPT}  A. Khrennikov (2025),   M. Ozawa,  F. Benninger,  O. Shor,  Coupling quantum-like cognition with the neuronal networks within generalized probability theory. Journal of Mathematical Psychology, 125, 102923.

\bibitem{K} A. N. Kolmogoroff. Grundbegriffe der Wahrscheinlichkeitsrechnung. Springer, Berlin (1933);  Kolmogorov, A. N.: Foundations of the Theory of Probability.  Chelsea Publ. Company, New York (1956).

\bibitem{Kriegel} Kriegel, U. (2020). The Oxford Handbook of the Philosophy of Consciousness. Oxford University Press.



\bibitem{Melkikh1} Melkikh, A. V., \& Sutormina, M. (2019). Protocells and LUCA: transport of substances from first physicochemical principles. Progress in Biophysics and Molecular Biology, 145, 85-104.
\bibitem{Melkikh2} Melkikh, A. V., \& Sutormina, M. I. (2022). From leaves to roots: Biophysical models of transport of substances in plants. Progress in Biophysics and Molecular Biology, 169, 53-83.


\bibitem{Ottinger1} H.C. \"Ottinger, Nonlinear thermodynamic quantum master equation: Properties and examples, Phys. Rev. A 82 (2010) 052119; 
\bibitem{Ottinger2} H.C. \"Ottinger, Stochastic process behind nonlinear thermodynamic quantum master equation, Phys. Rev. A 86 (2012) 032101.


\bibitem{Ozawa1} Ozawa M. Quantum measuring processes for continuous observables. J. Math. Phys., 25 (1984), 79-87

\bibitem{OzawaE}  Ozawa M., Khrennikov, A.  Application of theory of quantum instruments to psychology: Combination of question order effect with response replicability effect. Entropy, 22 (1) (2020), 37.1-9436

\bibitem{OzawaJMP} Ozawa M., Khrennikov A. Modeling combination of question order effect, response replicability effect, and QQ-equality with quantum instruments. J. Math. Psych., 100, 2021, 102491.

\bibitem{P} Penrose, R. (1989). \emph{The Emperor's new mind}, Oxford Univ. Press: New-York.

\bibitem{Pothos} Pothos, E.M.; Busemeyer, J.R. Quantum Cognition. Annu. Rev. Psychol. 2022, 73, 749–778.

\bibitem{Zeit}  A. Plotnitsky and E. Haven, The Quantum-Like Revolution. A Festschrift for Andrei Khrennikov with a foreword by 2022 Nobel Laureate Anton Zeilinger. Springer Nature, Berlin, 2023.

\bibitem{PLC} Plotnitsky, A. (2022). ``Most tantumising state of affairs'': Mathematical and non-mathematical in quantum-like understanding of thinking. Frontiers in Psychology, 13, 934776.

\bibitem{Singer1} Singer, W. (1999). Neuronal synchrony: A versatile code for the definition of relations? Neuron, 24, 49-65.

\bibitem{Singer2} Singer, W. (2018). The role of oscillations and synchrony in the development of the nervous system. Pages 15-32. In A. A. Benasich and U. Ribary, editors. Emergent Brain Dynamics. Prebirth to Adolescence. Str\"ungmann Forum Reports. MIT Press, Cambridge, MA. 

\bibitem{Singer3} Singer, W. (2024) The Mind-Body problem: Philosophical and neuroscientific attempts to bridge the gap between material and mental processes.  European Journal of Neuroscience,

\bibitem{Scholes1} Scholes G. D. (2024). Quantum-like states on complex synchronized networks. Proceedings of the Royal Society A, 480(2295), 20240209.


\bibitem{Shor1} Shor, O., Glik, A., Yaniv-Rosenfeld, A., Valevski, A., Weizman, A., Khrennikov, A., \& Benninger, F. (2021). EEG $p$-adic quantum potential accurately identifies depression, schizophrenia and cognitive decline. Plos one, 16(8), e0255529.

\bibitem{Patent}  Shor, O., Benninger, F., Khrennikov, A. EEG $P$-adic quantum potential in neuro-psychiatric diseases. US Provisional Patent Application No. 63/082,492.

\bibitem{Shor1_CHSH} Shor O, Benninger F, Khrennikov A. Dendrogramic Representation of Data: CHSH Violation vs. Nonergodicity. Entropy. 2021; 23(8):971

\bibitem{Shor2} Shor, O., Yaniv-Rosenfeld, A., Valevski, A., Weizman, A., Khrennikov, A., \& Benninger, F. (2023). EEG-based spatio-temporal relation signatures for the diagnosis of depression and schizophrenia. Sc. Rep. 13(1), 776.

\bibitem{TC} Takahashi, T., \& Cheon, T. (2012). A nonlinear neural population coding theory of quantum cognition and decision making. World Journal of Neuroscience, 2(4), 183-186.´

\bibitem{Tsekov} R. Tsekov A nonlinear master equation for open quantum systems. Fluct. Noise Lett. 20 (2021) 2130004.

\bibitem{V1} Vitiello, G. (1995). Dissipation and memory capacity in the quantum brain model, \emph{Int. J. Mod. Phys.}, B9,  973.
\bibitem{V2} Vitiello, G. (2001).  \emph{ My double unveiled: The dissipative quantum model of brain}, Advances in Consciousness Research,  John Benjamins Publishing Company. 
\bibitem{V3} Vallortigara, G.,  Vitiello, G. (2024). Brain asymmetry as minimization of free energy: a theoretical model. Royal Society Open Science, 11(7), 240465. 

\bibitem{VN} Von Neumann, J. {\it Mathematical Foundations of Quantum Mechanics}; Princeton University Press: Princeton, NJ, USA, 1955.

\bibitem{Wang} Wang Z., Busemeyer J.R. A quantum question order model supported by empirical tests of an a priori and precise prediction
Top. Cogn. Sci., 5 (2013), 689-710

\bibitem{Wang1} Wang Z., Solloway T., Shiffrin R.M., Busemeyer J.R.
Context effects produced by question orders reveal quantum nature of human judgments
Proc. Natl. Acad. Sci. USA, 111 (2014), 9431-9436  

\bibitem{Wi} Widdows, D., Rani, J., \& Pothos, E. M. (2023). Quantum circuit components for cognitive decision-making. Entropy, 25(4), 548.  


\end{thebibliography}
\end{document}